\documentclass[a4paper,11pt]{article}
\usepackage{jheppub}

\usepackage{comment}
\usepackage{amsmath,amssymb,extarrows,mathtools,graphicx,subfigure,setspace,bbold}
\usepackage{tikz}
\usetikzlibrary{decorations.pathreplacing}
\newcommand{\be}{\begin{equation}}
\newcommand{\bea}{\begin{eqnarray}}
\newcommand{\eea}{\end{eqnarray}}
\newcommand{\ba}{\begin{array}}
\newcommand{\ea}{\end{array}}
\newcommand{\bal}{\begin{align}}
\newcommand{\eal}{\end{align}}
\newcommand{\ee}{\end{equation}}
\newcommand{\bes}{\begin{equation*}}
\newcommand{\beas}{\begin{eqnarray*}}
\newcommand{\eeas}{\end{eqnarray*}}
\newcommand{\bas}{\begin{array*}}
\newcommand{\eas}{\end{array*}}
\newcommand{\ees}{\end{equation*}}

\title{Holographic quantum critical points in Lifshitz space-time}
\author{M. Reza Mohammadi Mozaffar,}
\author{Ali Mollabashi}

\affiliation{School of physics, Institute for Research in Fundamental Sciences (IPM)}
\emailAdd{m$_{-}$mohammadi@ipm.ir}
\emailAdd{mollabashi@ipm.ir}

\abstract{We study a minimally coupled charged scalar field in a charged Lifshitz background. For $z=2$, we find an analytic expression for the corresponding low energy retarded Green's function. Unlike the RN-AdS case, the position of the superfluid surfaces depends on the charge of the scalar field only through the IR scaling dimension. We show that by increasing the dynamical exponent, the dual theory becomes more stable. We also show that the background could suffer from an instability of the IR geometry leading to a bifurcating critical point. It also allows the existence of scalar hair, causing hybridized critical point. We have investigated stable and unstable regions in the parameter space.}
\keywords{Gauge-gravity correspondence, AdS-CFT Correspondence, Holography and condensed matter physics (AdS/CMT)}
\arxivnumber{1212.6635}

\begin{document}
\maketitle
\flushbottom
\section{Introduction}
The celebrated gauge/gravity duality \cite{Maldacena:1997re,Gubser:1998bc,W:1998} provides a useful tool to study strongly coupled quantum field theories via classical gravity. Over the past few years, this approach has been applied to study certain areas in condensed matter physics. In this context, one of the most challenging problems is to find a framework to study strongly correlated systems, such as non-Fermi liquids (for excellent reviews see \cite{Iqbal:2011ae,Hartnoll:2009sz}). An interesting feature of such systems is their behavior near quantum critical points (QCPs).

At the critical point, a physical system typically exhibits scaling symmetries which in general may be given by 
\be\label{scaling1}
t\rightarrow \lambda^zt,\;\;\;\;\;\;x_i\rightarrow \lambda x_i,
\ee
where $z$ is the dynamical exponent. Such a symmetry, known as Lifshitz scaling symmetry, is widely studied in condensed  matter physics starting with \cite{Hertz:1976zz}. The case of $z=1$ reduces to the relativistic case, while $z>1$ refers to anisotropic scaling symmetries.

To holographically study theories with this scaling symmetry, one may consider a classical gravity on the following background\cite{Kachru:2008yh}\footnote{As it has been mentioned in \cite{Kachru:2008yh}, although the metric is non-singular, it is not geodesically complete and has divergent tidal forces in the far infra-red. The situation where this divergence can be resolved is studied in \cite{Bao:2012yt,Harrison:2012vy}.}
\be\label{BacLif}
ds^2=L^2\left(- \frac{dt^2}{r^{2z}}+\frac{d\vec{x}^2}{r^2}+\frac{dr^2}{r^2}\right),
\ee
where $L$ is the radius of curvature. This metric is invariant under \eqref{scaling1} together with $r\to\lambda^{-1}r$. 
Also analytic solutions describing black-holes and black-branes in asymptotically Lifshitz spacetimes is constructed by considering Einstein-Maxwell-dilaton theories \cite{Tarrio:2011de}.

Following our previous work \cite{Alishahiha:2012nm}, where asymptotic Lifshitz solutions were probed by fermions, in this paper we will probe a charged Lifshitz black-brane by a charged scalar field in order to study the dual theory at the low energy limit. In this context, certain features of critical behavior, including bifurcating, hybridized, and marginal QCPs have been studied on RN-AdS background in four and five dimensions \cite{Iqbal:2011aj,Ren:2012hg}. Unlike thermal phase transitions, these quantum phase transitions could not be fully described by the Landau-Ginsburg paradigm. We will investigate these behaviors in a non-relativistic model.

We show that although this non-relativistic theory flows to an IR fixed point, the non-relativistic nature affects the critical behavior of the system via the dynamical exponent. Increasing the dynamical exponent suppresses the instabilities due to the violation of the BF bound in the near horizon AdS$_2$ region. We will focus on the case of $z=2$, which is claimed to be the margin of different physical behavior \cite{Tong:2012nf}. In this paper we will solve a scalar field analytically and work out the bosonic retarded Green's function at low frequency limit for standard quantization. The retarded Green's function has several zero modes, which could be a source of instability in the dual theory. The zero modes are located at
$$-\frac{1+\Delta_-}{2}+\nu_{k}=-n,$$ 
where $n$ is a non-negative integer, $\nu_k$ is the IR scaling dimension, and $\Delta_-$ is the scaling dimension of the dual operator defined by
\bea
\nu_k=\frac{1}{6}\sqrt{3m^2+3k^2+9-4e^2},\hspace{1cm}{\Delta_-}=-2+\sqrt{4+m^2}.
\eea
Unlike the case of RN-AdS$_5$, the locus of zero modes does not depend on the charge of the scalar field explicitly \cite{Ren:2012hg}.

We will proceed by studying different types of phase transitions in this model. The first type of phase transition occurs because of the violation of the BF bound deep in the space-time geometry. This point is called the "bifurcating" critical point. At such a critical point, the static susceptibility is not divergent but bifurcates into the complex plane. We will show that our model supports such a behavior and investigate different aspects of the phase diagram.

The second type of phase transition occurs when the system supports zero modes and leads to a hairy black-brane geometry. The onset of this behavior is known as the "hybridized" critical point. This critical point could be approached by turning on a double trace deformation. We study different regimes in the parameter space which support the hybridized critical point. We also show that only for $z>2$, the model could be deformed by a relevant or an irrelevant double trace operator, corresponding to the alternative and standard quantizations.

Approaching bifurcating and hybridized critical points at the same time leads to a third type of critical behavior. At such a point known as the "marginal" quantum critical point, we will study the Green's function which shows the behavior of marginal Fermi liquids.

This paper is organized as follows. In Sec. 2 we briefly review the charged Lifshitz black-brane solutions and their near horizon geometries. In Sec. 3 we shall discuss the non-relativistic behavior of the IR scaling dimension. Also the low energy retarded Green's function of a charged scalar field is calculated analytically for the case of $z=2$. In Sec. 4 we will study different types of critical behavior of the model. The last section is devoted to conclusions. The appendix contains some details of calculations.  

\section{Charged Lifshitz black-brane}
In order to study strongly correlated systems at finite density, one may consider a charged Lifshitz black-brane as a possible gravity dual. In the case where the temperature is much smaller than the chemical potential, the extremal limit of the black-brane must be considered. Thus we will consider an extremal charged Lifshitz black-brane in the bulk.

Pure Einstein gravity does not support Lifshitz geometry as a solution. One can obtain Lifshitz geometry either by coupling a massive gauge field, or by adding higher derivative terms to Einstein gravity \cite{Taylor:2008tg}. Asymptotic Lifshitz black-branes could be obtained as solutions of Einstein-Maxwell-dilaton theories\footnote{Superfluids and superconductors in which the gravitational theory includes a dilatonic field and are asymptotically AdS are studied in \cite{Salvio:2012at}.}. Consider the following action \cite{Tarrio:2011de}
\be\begin{split}
S&=\frac{1}{16\pi G_4}\int d^4x\sqrt{-g}\bigg[R-2\Lambda-\frac{1}{2}(\partial\phi)^2-
\frac{1}{4}e^{\lambda_1\phi} ({F^{(1)}})^2-\frac{1}{4} e^{\lambda_2\phi}({ F^{(2)}})^2\bigg]\\
&+\int d^3x\sqrt{-h}\;2K+S_{\mathrm{ct}}.
\end{split}
\ee
This model admits a charged black-brane solution where $z$ plays the role of critical exponent and $\Lambda, \lambda_1$ and $\lambda_2$ are given by
\be
\Lambda=-\frac{(z+1)(z+2)}{2L^2},\;\;\;\;\;\lambda_1=-\frac{2}{\sqrt{z-1}},
\;\;\;\;\;\lambda_2=\sqrt{z-1}.
\ee
with the following solution\footnote{Note that we have shifted the gauge fields by constants to make sure that $g^{\mu\nu}A^{(i)}_\mu A^{(i)}_\nu$ remains finite. We have set $L=1$ and also with a proper choice of the parameters the radius of horizon has also been set to unity.} 
\bea \label{1}
&&ds^2=-r^{2z}fdt^2+\frac{dr^2}{r^2f}+r^2d\vec{x}^2,\;\;\;\;\;\;
\;\;\;\;\;
f=1-\frac{1+r_0^{2(z+1)}}{r^{z+2}}+\frac{r_0^{2(z+1)}}{r^{2(z+1)}},
\cr
&&e^{\sqrt{z-1}\phi}=
\frac{\kappa^2 }{4zr_0^{2(z+1)}}r^{2(z-1)},\cr
&&
A^{(1)}_t=-\mu^{(1)}\bigg(1-r^{2+z}\bigg),\;\;\;\;
A^{(2)}_t=\mu^{(2)}\bigg(1-\frac{1}{r^{z}}\bigg),
\eea
where the mass and charge of the black-brane is determined in terms of $r_0$ and $\kappa$, and
\be \label{2}
\mu^{(1)}=\sqrt{\frac{2(z-1)}{z+2}}\left(\frac{\kappa^2} {4zr_0^{2(z+1)}}\right)^{\frac{1}{z-1}},\;\;\;\;\;\;\;\;\;\;\;\;
\mu^{(2)}=\frac{4r_0^{2(z+1)}}{\kappa }.
\ee
Although $A^{(1)}$ diverges near the boundary, the on-shell action becomes finite by choosing
\be\begin{split}
S_{\mathrm{ct}} &= -\int d^3x\sqrt{-h}\Big[2(z+1)-\frac{z+2}{2}(\alpha-\alpha_0)\\
&\hspace{4cm}+ \frac{(z+2)^2}{16(z^2-1)}\left(3+2z+\sqrt{(z+2)(10+9z)}\right)(\alpha-\alpha_0)^2\Big]
\end{split}
\ee
where $\alpha=e^{\lambda\phi}h^{ab}A_aA_b$ and $\alpha_0=-\frac{2(z-1)}{(z+2)}$ is its background value\footnote{For the case where Lifshitz solution is constructed by a massive gauge field, the renormalized action is constructed in \cite{Ross:2011gu,Baggio:2011cp,Mann:2011hg}.}. Also one can define a well defined 'complex' stress energy tensor for such theories (see \cite{Ross:2009ar, Ross:2011gu}). 

Since $A^{(1)}$ diverges at the boundary, it can not be treated as a chemical potential of boundary theory and we are forced to couple matter fields to $A^{(2)}$.

The Hawking temperature reads
\be
T=\frac{z+2}{4\pi}\left(1-\frac{z}{z+2}r_0^{2(z+1)}\right).
\ee
Using the general idea of AdS/CFT correspondence, the physics at low energy is governed by the near horizon modes. Thus  we will study a charged scalar on the near horizon background. At zero temperature where
$r_0^{2(z+1)}=\frac{z+2}{z}$ and $f=1-2(1+z^{-1})r^{-(z+2)}+(1+2z^{-1})r^{-2(z+1)}$, setting
\be \label{scaling}
r-1=\frac{\epsilon }{(z+1)(z+2) \xi},\;\;\;\;\;\;\;\;\;t=\frac{1}{\epsilon} \tau,
\ee
the near horizon background can be obtained by taking the limit $\epsilon\rightarrow 0$ where one finds
\bea\label{AdS2}
ds^2=\frac{l_z^2}{\xi^2}\left(-d\tau^2+d\xi^2\right)+ dx_2^2
\eea
and $l_z^{-2}=(z+1)(z+2)$. In this limit the dilaton and gauge fields reads
\bea
e^{\sqrt{z-1}\phi}=\frac{\kappa^2}{4(z+2) },\hspace{1cm}
A^{(1)}_\tau=\frac{\mu^{(1)} }{(z+1)\xi},\hspace{1cm}A^{(2)}_\tau=\frac{4}{(z+1)\kappa\xi}.
\eea
The near horizon geometry takes the form of AdS$_2\times \mathbb{R}^2$ and is  invariant under the following scale transformations
\begin{eqnarray}
\tau \rightarrow \lambda \tau,\hspace{1cm}\xi \rightarrow \lambda \xi,\hspace{1cm} x_i \rightarrow x_i.
\end{eqnarray} 
Therefore the low energy physics is described by an emergent IR CFT. As a result, we would expect the model to exhibit superfluid surfaces whose physics is governed by an IR fixed point. The $\mathbb{R}^2$ factor of the near horizon geometry together with the AdS$_2$ factor leads to the so-called semi-local quantum liquid (SLQL) phase \cite{Faulkner:2009wj}.
\section{Analytic Green's function}
In order to find the retarded Green's function one must solve the scalar field equation of motion on background \eqref{1}, which reads  
\bea\label{scalar}
r^{-z-1}\partial_r \left(r^{z+3}f \partial_r \phi\right)+\left[\frac{(\omega+qA_t^{(2)})^2}{r^{2z}f}-m^2-\frac{k^2}{r^2}\right]\phi=0.
\eea
In the low frequency limit, since $f$ has a double zero at the horizon, one can not take the $\omega\to0$ limit strait forwardly. To treat this regime, following \cite{Faulkner:2009wj}, the geometry may be divided into IR and UV regions. In the IR region one can not neglect the $\omega$ term due to the singularity of $f$ near the horizon. But in the UV region where $f$ is regular, one can set $\omega=0$ at first order. Matching the solutions in the intersection region could lead to the retarded Green's function of the whole geometry at leading order in $\omega$. 

\begin{figure}
\centering
\begin{subfigure}
\centering
\includegraphics[scale=.7]{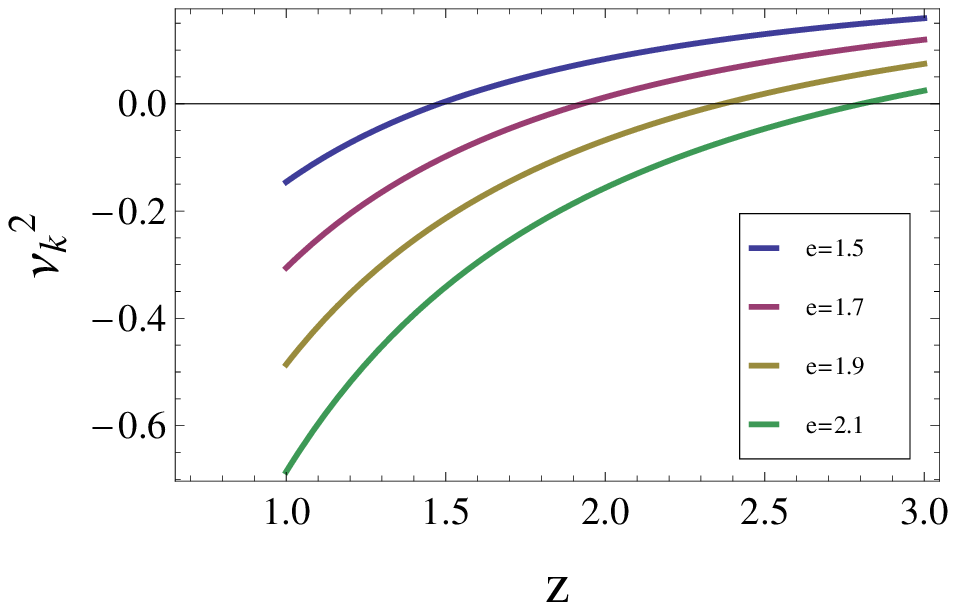}
\end{subfigure}
\begin{subfigure}
\centering
\includegraphics[scale=.7]{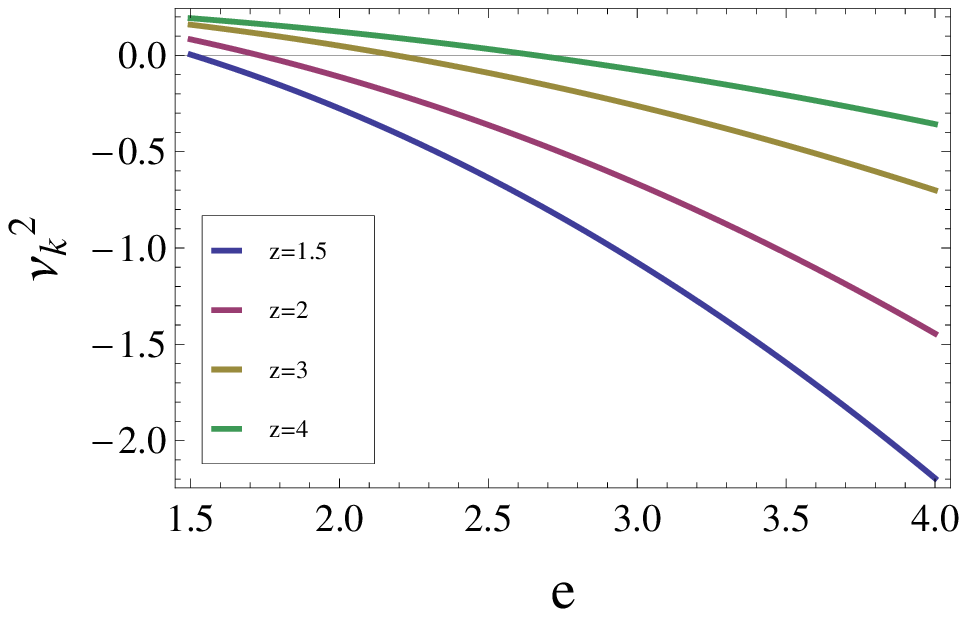}
\end{subfigure}
\caption{\textit{Left plot}: Square of $\nu_k$ as function of dynamical exponent is shown for different values of charge. As the dynamical exponent increases, the systems tends to IR stable region. The absolute value of charge compensates reaching the stable region. \textit{Right plot}: Square of $\nu_k$ as function of charge for different values of $z$. As the charge increases, $\nu_k$ tends to zero and finally takes pure imaginary values which causes IR instability. Notice that for larger dynamical exponents the systems reaches the unstable region softer. In both plots $m=1$ and $k=0$.}
\label{nu}
\end{figure}

\subsection{Low energy behavior and IR Green's function}
Solving Eq. \eqref{scalar} in the IR geometry \eqref{AdS2} leads to the following IR Green's function (for details see the appendix)
\begin{eqnarray}\label{IRGreen}
\mathcal{G}_k(\omega)=e^{-i\pi \nu_k}\frac{\Gamma(-2\nu_k)}{\Gamma(2\nu_k)}\frac{\Gamma(\frac{1}{2}-ie_z+\nu_k)}{\Gamma(\frac{1}{2}-ie_z-\nu_k)}(2\omega)^{2\nu_k},
\end{eqnarray}
where
\bea
\nu_k=\sqrt{\frac{1}{4}+\frac{m^2+k^2}{(z+1)(z+2)}-e_z^2}\;,
\eea
and $e_z=\frac{4q}{(z+1)\kappa}$. The result is similar to the RN-AdS case, where the IR Green's function scales with frequency for arbitrary spatial momenta.

Mixing between the scalar and the gauge field leads to a negative effective mass squared for the scalar field. So in a region of the parameter space, $\nu_k$ can become imaginary for large enough values of $e_z$ that results in possible scalar pair productions and thus instability of the IR geometry. This process is similar to what happens in the RN-AdS black-brane case \cite{Iqbal:2011in,Iqbal:2011aj,Iqbal:2011ae}. In contrast with this case in which the gauge field couples to matter fields in the near horizon geometry, there exists some examples where the gauge field vanishes in the near horizon limit, so there is no pair production and thus the geometry remains stable (an example is studied in \cite{Gubser:2012yb,Alishahiha:2012ad}).

The $z$ dependence of $\nu_k$ shows that the net effect of increasing the dynamical exponent pushes the system toward more stable regions 
(see Fig. \ref{nu})
\footnote{One can see from \eqref{2} that for $z<1$, the gauge field becomes complex, so we always consider $z>1$.}. Also notice that for large values of $z$, for arbitrary mass and charge, the system becomes stable. 
\subsection{UV region and matching}
In order to match the solutions of the inner and outer regions at leading order in $\omega$-expansion, we have to solve  the Klein-Gordon equation \eqref{EOM1} in the outer region at $\omega=0$ which reads\footnote{From now on we will drop the ${(0)}$ subscript which indicates the $\omega$-expansion leading order.}
\begin{eqnarray}\label{UVleading}
r^{-z-1}\partial_r \left(r^{z+3}f \partial_r \phi_{(0)}\right)+\left[\frac{(z+2)^2e^2}{z^2r^{2z}f}\left(1-\frac{1}{r^z}\right)^2-m^2-\frac{k^2}{r^2}\right]\phi_{(0)}=0.
\end{eqnarray}
In what follows we will restrict ourselves on the $z=2$ case where the above equation could be solved exactly.
Near the boundary ($r \to \infty$) the solution behaves as
\begin{eqnarray}\label{z=2infty}
\phi(r \rightarrow \infty)=A\,r^{\Delta_-}+B\,r^{\Delta_+}, \hspace{1cm}{\Delta_\pm}=-2\mp\nu_U=-2\mp\sqrt{4+m^2},
\end{eqnarray}
where the coefficients $A$ and $B$ represent the source and response functions respectively. The scaling dimension of the dual operator is real-valued when $m^2>-4$, where the vacuum of the dual theory could be stable \cite{Kachru:2008yh}. This bound is called the generalized $z$-dependent Breitenlohner-Freedman bound which in a $d+1$-dimensional Lifshitz space-time takes the following general form \cite{Giacomini:2012hg}
\begin{figure}
\begin{center}
\begin{tikzpicture}[scale=1]
\draw[->](-7.5,9.5) -- (-1,9.5) node[anchor=west]{$m^2$};
\draw[semithick] (-7,9.4) -- (-7,9.6);
\draw[semithick] (-2,9.4) -- (-2,9.6);
\draw[semithick] (-4,9.4) -- (-4,9.6);
\draw (-7,8.9) node {$m^2_{\mathrm{BF}}$};
\draw (-4,8.8) node {$-\frac{(z+1)(z+2)}{4}$};
\draw (-2,8.9) node {$m^2_{\mathrm{BF}}+1$};
\draw (-4,7.5) node {$z<2$};
\draw [thick, decorate,decoration={brace,amplitude=10pt},xshift=0pt,yshift=2pt] (-7,9.7) -- (-2,9.7) node [black,midway,yshift=16pt] {\footnotesize alternative quantization allowed};
\draw [thick, decorate,decoration={brace,amplitude=7pt,mirror,raise=2pt},xshift=0pt,yshift=0pt] (-7,9.4) -- (-4,9.4) node [black,midway,yshift=-28pt] {\footnotesize Oscillatory region};
\draw[->] (-5.5,9) -- (-5.5,8.6);
\draw[->](1.5,9.5) -- (8,9.5) node[anchor=west]{$m^2$};
\draw[semithick] (2,9.4) -- (2,9.6);
\draw[semithick] (5,9.4) -- (5,9.6);
\draw[semithick] (7,9.4) -- (7,9.6);
\draw (2,9.17) node {$m^2_{\mathrm{BF}}$};
\draw (7,9.1) node {$-\frac{(z+1)(z+2)}{4}$};
\draw (5,9.2) node {$m^2_{\mathrm{BF}}+1$};
\draw (5,7.5) node {$z>2$};
\draw [thick, decorate,decoration={brace,amplitude=10pt},xshift=0pt,yshift=2pt] (2,9.7) -- (5,9.7) node [black,midway,yshift=16pt] {\footnotesize alternative quantization allowed};
\draw [thick, decorate,decoration={brace,amplitude=7pt,mirror,raise=2pt},xshift=0pt,yshift=0pt] (2,9) -- (7,9) node [black,midway,yshift=-20pt] {\footnotesize Oscillatory region};
\draw[->](-3,5.5) -- (3.5,5.5) node[anchor=west]{$m^2$};
\draw[semithick] (-2.5,5.4) -- (-2.5,5.6);
\draw[semithick] (3,5.4) -- (3,5.6);
\draw (-2.5,5.2) node {$-4$};
\draw (3,5.2) node {$-3$};
\draw (.3,3.5) node {$z=2$};
\draw [thick, decorate,decoration={brace,amplitude=10pt},xshift=0pt,yshift=2pt] (-2.5,5.7) -- (3,5.7) node [black,midway,yshift=16pt] {\footnotesize alternative quantization allowed};
\draw [thick, decorate,decoration={brace,amplitude=7pt,mirror,raise=2pt},xshift=0pt,yshift=0pt] (-2.5,5.1) -- (3,5.1) node [black,midway,yshift=-20pt] {\footnotesize Oscillatory region};
\end{tikzpicture}
\caption{Alternative quantization region versus oscillatory region for Lifshitz$_4$ and general dynamical exponent. For the case of $z<2$ alternative quantization is allowed for $-\frac{(z+1)(z+2)}{4}<m^2<m_{\mathrm{BF}}^2+1$. This case coincides with the case of RN-AdS$_4$ for $z=1$. For the case of $z=2$ the oscillatory region coincides with the region where alternative quantization is allowed thus there is no room for alternative quantization. This case coincides with the case of RN-AdS$_5$. In the case of $z>2$ the region where alternative quantization is allowed is a subregion of the oscillatory region thus again there is no room for alternative quantization.}
\label{Lif4}
\end{center}
\end{figure}
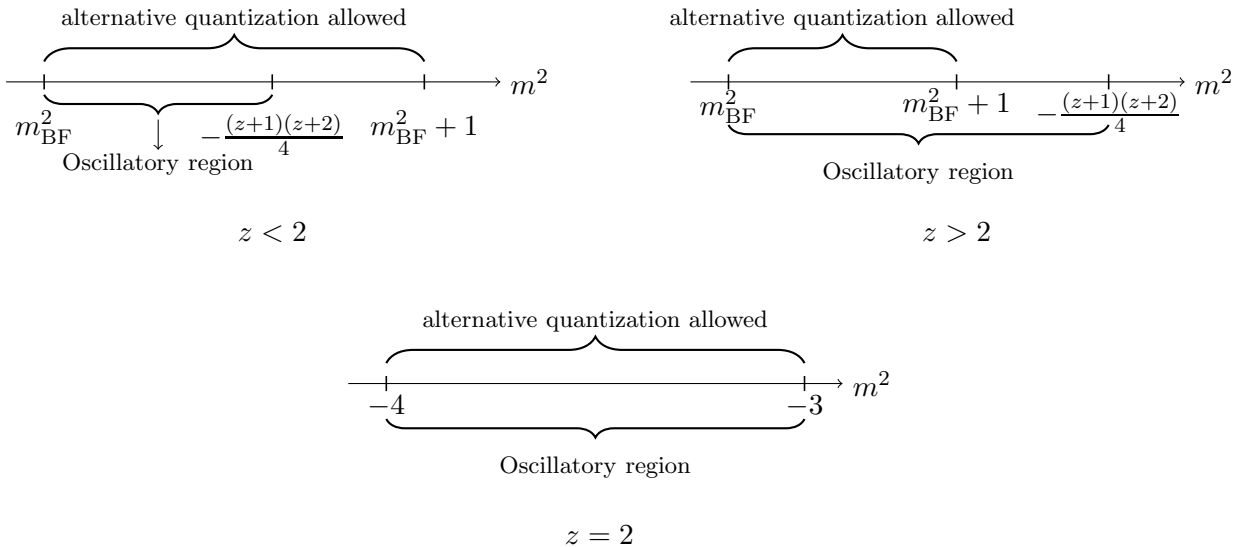
\begin{eqnarray}\label{BFLif}
m^2>m^2_{\mathrm{BF}};\hspace{2cm}m^2_{\mathrm{BF}}=-\frac{(d+z-1)^2}{4L^2}.
\end{eqnarray}
In the case of AdS$_{d+1}$ this bound is $m^2>-\frac{d^2}{4L^2}$,
thus the bound is less restrictive on Lifshitz geometries compared to AdS ones. In both types of geometries, the bound on the whole space-time does not guarantee the stability of the IR region which contains an AdS$_2$ factor. In our case the BF bound on the near horizon geometry\footnote{Remember that our AdS$_2$ radius is $l_2^{-2}=12$.} is $m^2>-3$ while the whole geometry is stable for $m^2>-4$. Thus there exists a window ($-4<m^2<-3$) that the four dimensional space is stable while its near horizon geometry becomes unstable.

The alternative quantization, referred to as Neuman boundary condition is also allowed  in the range \cite{Keeler:2012mb, Andrade:2012xy}
\be\label{NW}
m^2_{\mathrm{BF}}<m^2<m^2_{\mathrm{BF}}+z^2.
\ee

Although the normalizability seems to allow masses in the above range for alternative quantization, a more detailed study which was carried by \cite{Andrade:2012xy} reveals that there are severe instabilities unless
\be
m^2_{BF} < m^2 < m^2_{BF} + 1.
\ee
As discussed in \cite{Andrade:2012xy}, this is essentially a UV effect in the field theory and this instability will affect any solution which asymptotically approaches Lifshitz solution with the Neumann boundary condition\footnote{We kindly thank the authors of \cite{Andrade:2012xy} to bring our attention to this effect.}.

The exact solution of Eq. \eqref{UVleading} is
\begin{eqnarray}\label{exactz=2}
\phi(r)&=&(r^2-1)^{\nu_k-\frac{1}{2}}\Bigg[c_1(\frac{r^2+2}{3})^{\frac{-ie}{3}}F\left(\nu_k-\frac{1+\Delta_+}{2}-\frac{ie}{3},\nu_k-\frac{1+\Delta_-}{2}-\frac{ie}{3},1-\frac{2ie}{3},\frac{r^2+2}{3}\right)\nonumber\\&&
+c_2(\frac{r^2+2}{3})^{\frac{ie}{3}}F\left(\nu_k-\frac{1+\Delta_+}{2}+\frac{ie}{3},\nu_k-\frac{1+\Delta_-}{2}+\frac{ie}{3},1+\frac{2ie}{3},\frac{r^2+2}{3}\right)\Bigg]
\end{eqnarray}
where
\begin{eqnarray}\label{nuk}
\nu_k=\frac{1}{6}\sqrt{3m^2+3k^2+9-4e^2}.
\end{eqnarray}
Expanding this solution in the near horizon limit and by using Eq. \eqref{near matching}, one can read $c_1$ and $c_2$. Considering the near boundary limit of the solution \eqref{exactz=2}, one can read the retarded Green's function of the full geometry at leading order in $\omega$. The retarded Green's function takes the form of the well-known master formula of \cite{Faulkner:2009wj} which is
\begin{eqnarray}\label{master}
G_R(\omega,k)=\frac{b_+^{(0)}(k)+b_-^{(0)}(k)\mathcal{G}_k(\omega)}{a_+^{(0)}(k)+a_-^{(0)}(k)\mathcal{G}_k(\omega)},
\end{eqnarray}
where in our case we find
\begin{eqnarray}\label{coef}
\begin{pmatrix}
a_+^{(0)}&a_-^{(0)}\\b_+^{(0)}&b_-^{(0)}
\end{pmatrix}=\frac{\nu_k}{\nu_U}
\begin{pmatrix}
\frac{2^{\frac{1}{2}+\nu_k}3^{2\nu_k-\frac{\nu_U}{2}}\Gamma(2\nu_k)\Gamma(-1-\Delta_+)}{\Gamma(-\frac{1+\Delta_+}{2}+\nu_k-\frac{ie}{3})\Gamma(-\frac{1+\Delta_+}{2}+\nu_k+\frac{ie}{3})}&\frac{-2^{\frac{1}{2}-\nu_k}3^{-2\nu_k-\frac{\nu_U}{2}}\Gamma(-2\nu_k)\Gamma(-1-\Delta_+)}{\Gamma(-\frac{1+\Delta_+}{2}-\nu_k-\frac{ie}{3})\Gamma(-\frac{1+\Delta_+}{2}-\nu_k+\frac{ie}{3})}\\\frac{-2^{\frac{1}{2}+\nu_k}3^{2\nu_k+\frac{\nu_U}{2}}\Gamma(2\nu_k)\Gamma(-1-\Delta_-)}{\Gamma(-\frac{1+\Delta_-}{2}+\nu_k-\frac{ie}{3})\Gamma(-\frac{1+\Delta_-}{2}+\nu_k+\frac{ie}{3})}&\frac{2^{\frac{1}{2}-\nu_k}3^{-2\nu_k+\frac{\nu_U}{2}}\Gamma(-2\nu_k)\Gamma(-1-\Delta_-)}{\Gamma(-\frac{1+\Delta_-}{2}-\nu_k-\frac{ie}{3})\Gamma(-\frac{1+\Delta_-}{2}-\nu_k+\frac{ie}{3})}
\end{pmatrix}.
\end{eqnarray}
One can easily check that the determinant of the above matrix is equal to $\nu_k/\nu_U$, which is a consequence of equality of the Wronskians in the infinity and the horizon of \eqref{UVleading}. As long as $\nu_k$ is real-valued, the boundary condition of the scalar field in the UV region and thus these coefficients become real.
By expanding Eq. \eqref{master} near $\omega=0$ one finds
\begin{eqnarray}\label{GRo}
G_R(\omega \rightarrow0,k)=\frac{b_+^{(0)}(k)}{a_+^{(0)}(k)}+\frac{\nu_k}{\nu_U}\frac{\mathcal{G}_k(\omega)}{a_+^{(0)2}(k)}.
\end{eqnarray}
In the case of real IR scaling dimension, the spectral function reads
\begin{eqnarray}
\mathcal{A}(\omega,k)=\frac{\nu_k}{\nu_U}\frac{{\rm Im}\mathcal{G}_k(\omega)}{a_+^{(0)2}(k)},
\end{eqnarray}
thus the near horizon fluctuations are dominant in the low energy regime and the static susceptibility vanishes. In the case of complex-valued IR scaling dimension, the matching coefficients become complex and thus the static susceptibility bifurcates into the complex plane \cite{Iqbal:2011aj}.   

\section{Instabilities, phase transitions and QCPs}
In this section we investigate the analytic behavior of the low energy Green's function which can be used to find zero modes (locus of Fermi surfaces) and explore the existence of quasi particles as excitations around it. Also one can extract information about the possible quantum phase transitions and its QCPs. In what follows we will study the QCPs mentioned in Sec. 1 and try to explain their essences in our model.
\subsection{Zero modes and phase diagrams}
In order to find zero modes of a bosonic system, we must find the poles of the retarded Green's function at zero frequency.
Following the terminology of \cite{Ren:2012hg}, we will refer to the location of these superfluid surfaces by $k_S$
similar to the location of Fermi surfaces, $k_F$, in fermionic systems. A simple analysis shows that the poles are due to the divergences of $b_+^{(0)}(k)$. After some gymnastics with the Gamma functions one finds that the superfluid surfaces are located at\footnote{We have used the following identity\cite{Arfken}
\begin{eqnarray*}
\left|\Gamma(\alpha+i\beta)\right|=\left|\Gamma(\alpha)\right| \prod_{n=0}^\infty \left(1+\frac{\beta^2}{(\alpha+n)^2}\right)^{-\frac{1}{2}},
\end{eqnarray*}where $\alpha$ and $\beta$ are real functions.}
\begin{eqnarray*}
-\frac{1+\Delta_-}{2}+\nu_{k}=-n,
\end{eqnarray*}
where $n$ is a non-negative integer. Note that unlike the case of extremal RN-AdS$_5$ black-brane \cite{Ren:2012hg}, the position of superfluid surfaces does not depend on the charge of scalar field explicitly, although it depends implicitly through the IR scaling dimension.
The necessary condition for $k_S$ to be real requires $\nu_k\ge0$ which implies
\begin{eqnarray}\label{mbound}
m^2 \ge(2n+1)^2-4.
\end{eqnarray}
This condition shows that all the zero modes are above the generalized BF bound \eqref{BFLif}. In the $n=0$ case
\begin{eqnarray}\label{ks}
k_S^2=3(1+\Delta_-)^2+\frac{4e^2}{3}-m^2-3,
\end{eqnarray}
where $k_S$ vanishes for $e=0$ and $m^2=-3$.

\begin{figure}
\centering
\begin{subfigure}
\centering
\includegraphics[scale=.48]{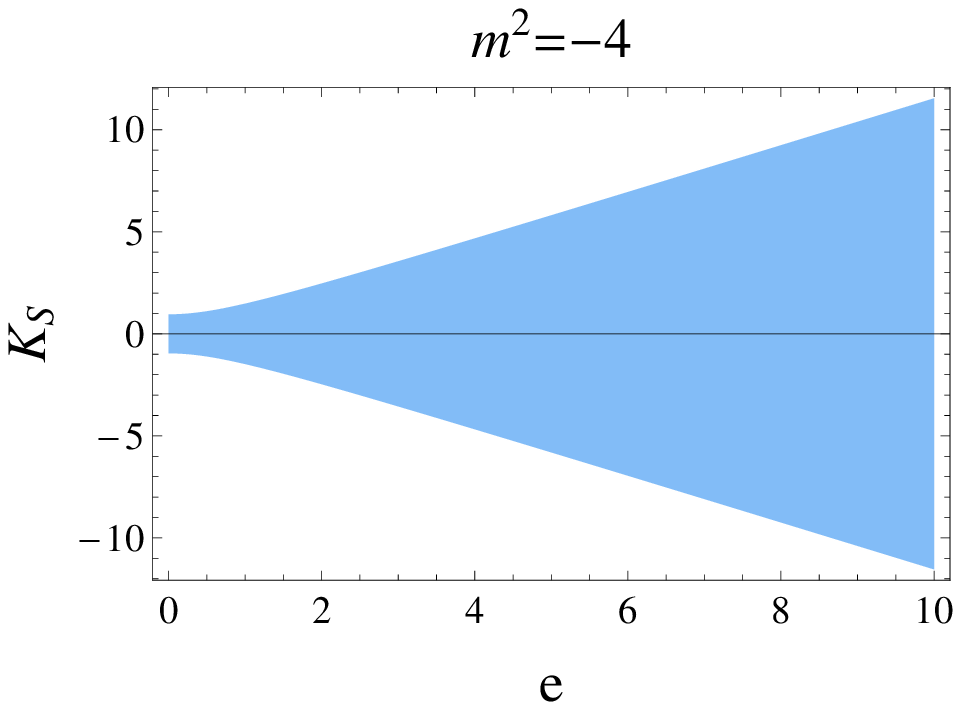}
\end{subfigure}
\begin{subfigure}
\centering
\includegraphics[scale=.48]{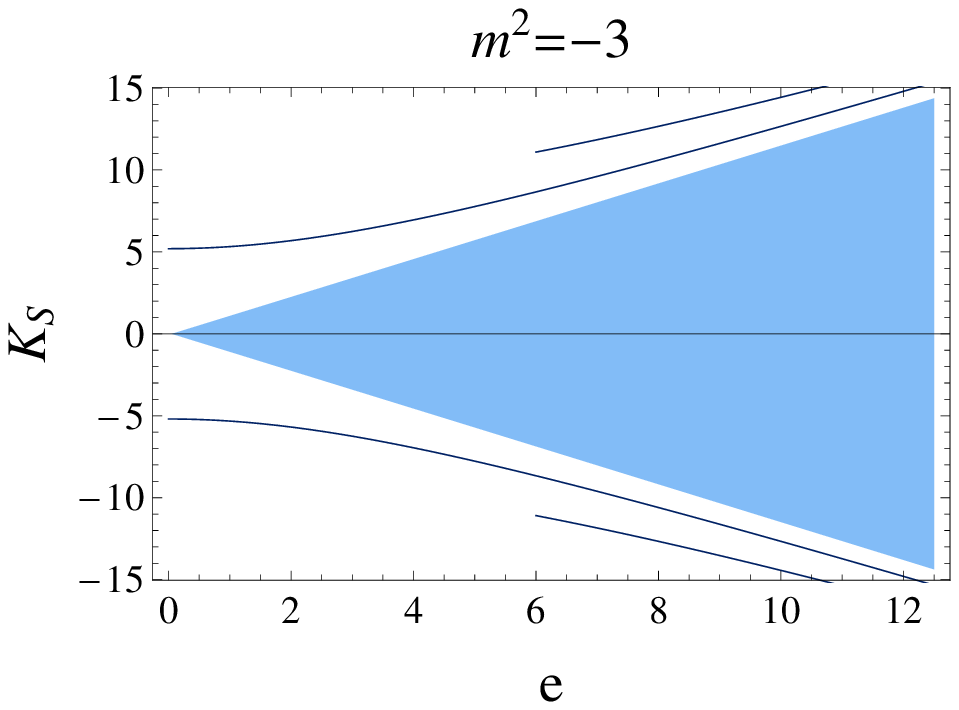}
\end{subfigure}
\begin{subfigure}
\centering
\includegraphics[scale=.48]{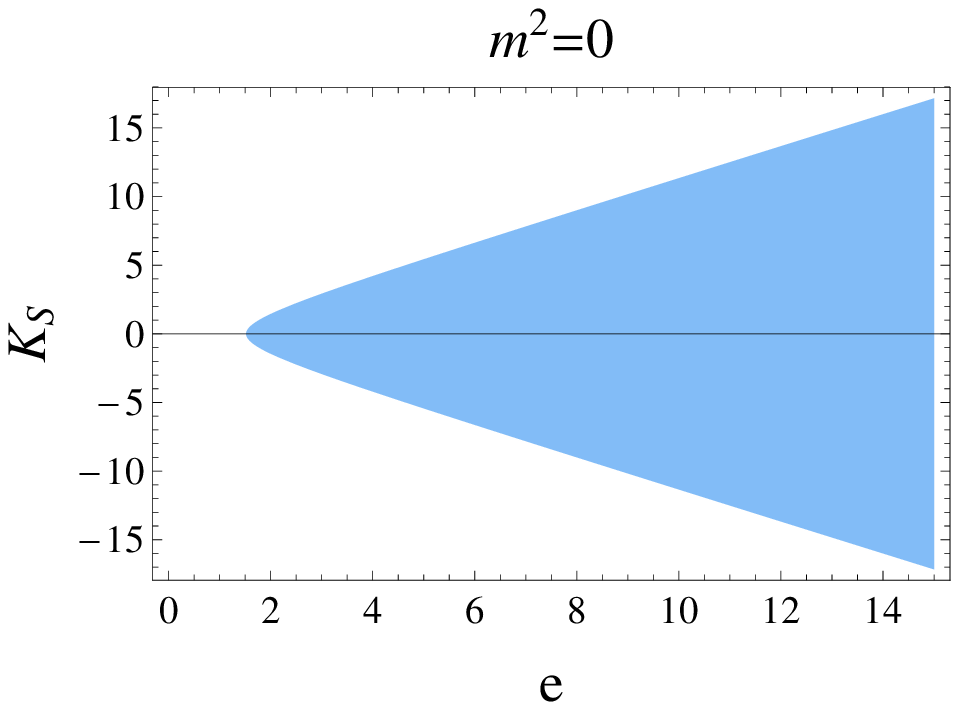}
\end{subfigure}
\caption{Phase diagrams of a charged scalar in Lifshitz background ($z=2$) with standard quantization for different mass parameters. The solid lines denote the zero modes while the blue region corresponds to the oscillatory region.}
\label{KsQ}
\end{figure}

The critical value for momentum where the IR scaling dimension becomes imaginary is 
\begin{eqnarray}\label{kcrit}
k_{c}^2=\frac{4e^2}{3}-m^2-3,
\end{eqnarray}
where
\begin{eqnarray}\label{e}
e^2\ge \frac{3}{4}(m^2+3).
\end{eqnarray}
Note that we always have $k_S>k_{c}$ and also remember that $k<k_{c}$ denotes the oscillatory region \cite{Faulkner:2009wj}. Outside the oscillatory region, zero modes must satisfy both conditions described in Eqs. \eqref{mbound} and \eqref{e} which leads to
\begin{eqnarray}
e^2\ge 3(n+\frac{1}{2})^2-\frac{3}{4}.
\end{eqnarray}
These constraints enlightens some features of the parameter space which is shown in Fig. \ref{KsQ}. Note that according to Eq. \eqref{z=2infty}, we always consider $m^2\ge-4$, to ensure that the vacuum of the dual field theory is stable.
If we deform the system slightly near its zero modes some interesting behavior could occur. If the system supports poles in both the upper and lower half of the complex $\omega$-plane, it is a sign of instability. In order to further study the stability conditions of the system we have focused on the poles in the complex $\omega$-plane. To do so we have solved the scalar field retarded Green's function numerically to all orders in frequency. This could be done by imposing the near boundary solution of the IR region as boundary conditions of the UV region. The system supports several poles for different set of parameters. For a given mass parameter, at $e=0$ and $k=0$, the poles lie in the lower half plan while increasing $e$ pushes them upstairs. For a given mass parameter and a large enough $e$, the schematic result is shown in Fig. \ref{COP}. As $|k|$ increases, the poles approach the normal mode (origin) and by further increasing $|k|$, they finally jump into the lower half plan. \subsection{Bifurcating quantum critical point}
When the BF bound of the near horizon geometry is violated, the Green's function bifurcates into the complex $\omega$-plane, known as bifurcating quantum phase transition. At this critical point the effective AdS$_2$ mass vanishes at zero spatial momentum\footnote{We set $k=0$, in order to neglect the effect of the extra $\mathbb{R}^2$ factor.}. At such a critical point the zero frequency Green's function does not diverge but bifurcates into the complex plane. The effective AdS$_2$ mass could be tuned by the value of the scalar charge. Defining
\begin{eqnarray*}
\nu=\frac{1}{6}\sqrt{3m^2+9-4e^2}\equiv\sqrt{u},
\end{eqnarray*}
a bifurcating QCP happens at
\begin{eqnarray*}
u=u_c=0.
\end{eqnarray*}
Near $\nu=0$, the matching matrix could be expanded
\begin{eqnarray*}
\begin{pmatrix}
a_+^{(0)}&a_-^{(0)}\\b_+^{(0)}&b_-^{(0)}
\end{pmatrix}(\nu\rightarrow0)=\begin{pmatrix}
\alpha&\alpha\\
\beta&\beta
\end{pmatrix}+\nu
\begin{pmatrix}
\tilde{\alpha}&\tilde{\alpha}\\
\tilde{\beta}&\tilde{\beta}
\end{pmatrix},
\end{eqnarray*}
where
\begin{eqnarray*}
\alpha&=&\frac{3^{-\frac{\nu_U}{2}}\Gamma(\nu_U)}{\sqrt{2}\Gamma\left(-\frac{1+\Delta_+}{2}-\frac{ie}{3}\right)\Gamma\left(-\frac{1+\Delta_+}{2}+\frac{ie}{3}\right)},\nonumber\\
\tilde{\alpha}&=&-\frac{3^{-\frac{\nu_U}{2}}\Gamma(\nu_U)\left[\psi\left(-\frac{1+\Delta_+}{2}-\frac{ie}{3}\right)+\psi\left(-\frac{1+\Delta_+}{2}+\frac{ie}{3}\right)+2\gamma-\ln18\right]}{\sqrt{2}\Gamma\left(-\frac{1+\Delta_+}{2}-\frac{ie}{3}\right)\Gamma\left(-\frac{1+\Delta_+}{2}+\frac{ie}{3}\right)},\nonumber\\
\beta&=&\frac{3^{\frac{\nu_U}{2}}\Gamma(-\nu_U)}{\sqrt{2}\Gamma\left(-\frac{1+\Delta_-}{2}-\frac{ie}{3}\right)\Gamma\left(-\frac{1+\Delta_-}{2}+\frac{ie}{3}\right)},\nonumber\\
\tilde{\beta}&=&-\frac{3^{-\frac{\nu_U}{2}}\Gamma(-\nu_U)\left[\psi\left(-\frac{1+\Delta_-}{2}-\frac{ie}{3}\right)+\psi\left(-\frac{1+\Delta_-}{2}+\frac{ie}{3}\right)+2\gamma-\ln18\right]}{\sqrt{2}\Gamma\left(-\frac{1+\Delta_-}{2}-\frac{ie}{3}\right)\Gamma\left(-\frac{1+\Delta_-}{2}+\frac{ie}{3}\right)},
\end{eqnarray*}
and $\psi(x)$ denotes the digamma function. In this case the determinant condition implies that $\alpha\tilde{\beta}-\beta\tilde{\alpha}=\frac{-1}{2\nu_U}$. Expanding the IR Green's function near the bifurcating critical point leads to
\bea\label{Gbi}
\mathcal{G}_{k=0}(\omega)&=&-1+2\nu\mathcal{G}_0(\omega)\nonumber\\
\mathcal{G}_0(\omega)&=&-2\gamma-\psi\left(\frac{1}{2}-\frac{ie}{3}\right)-\ln(-2i\omega).
\eea
\begin{figure}
\centering
\includegraphics[scale=.7]{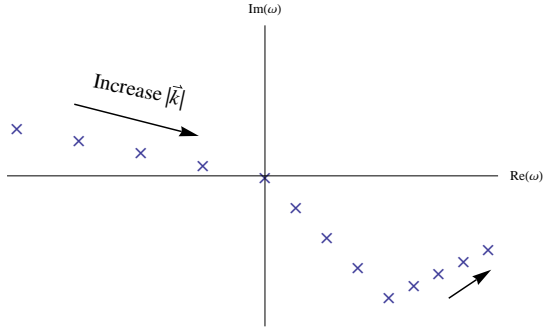}
\caption{Schematic plot of the poles of the Green's function for large scalar charge $e$. This bahavior is found by solving the scalar field equation of motion numerically in the UV region, using the near boundary solution of the IR region as boundary condition. These poles correspond to the condensed region ($u<0$) and the most right pole refers to $u=0$.}
\label{COP}
\end{figure}
Thus the Green's function for the whole geometry near such a critical point takes the form of
\begin{eqnarray*}
G=\frac{\beta\mathcal{G}_0(\omega)+\tilde{\beta}}{\alpha\mathcal{G}_0(\omega)+\tilde{\alpha}}.
\end{eqnarray*}
Also one can find similar results for the finite temperature case by using $\mathcal{G}_k^T$ from Eq. \eqref{scalarIRGreenTemp}. In this case one finds
\bea\label{GbiT}
\mathcal{G}^T_{k=0}(\omega)&=&-1+2\nu\mathcal{G}^T_0(\omega),\nonumber\\
\mathcal{G}^T_0(\omega)&=&\mathcal{G}_0(\omega)-\psi\left(\frac{1}{2}+\frac{ie}{3}-\frac{i\omega}{2\pi T}\right)+\ln\left(\frac{-i\omega}{2\pi T}\right).
\eea
In the neutral case it reduces to
\bea
\mathcal{G}^T_0(\omega)=-\gamma-\psi\left(\frac{1}{2}-\frac{i\omega}{2\pi T}\right)-\ln\left(\pi T\right).
\eea
As it could be seen in Fig. \ref{COP}, in the condensed side ($u<0$), the system supports infinite number of poles in the upper half $\omega$-plane. This is similar to the case of RN-AdS black-brane \cite{Iqbal:2011aj}. 
\subsection{Hybridiezed quantum critical point}
One way to achieve the zero modes (at $k=0$) is to add a double trace deformation to the CFT \cite{Witten:2001ua}\footnote{It could also be achieved by considering superfluid velocity.}. Under double trace deformations, the dual theory flows into a new CFT where the scaling dimension of $\mathcal{O}$ changes from $\Delta$ to $d-\Delta$. As a result of adding a double trace operator
\begin{eqnarray*}
\frac{\kappa_+}{2}\int{d^dx\;\mathcal{O}^2},
\end{eqnarray*}
to the action, the Green's function in the momentum space becomes a geometric sum over connected diagrams. In this case the Green's function deforms as
\begin{eqnarray*}
G^{(\kappa)}_R(\omega,k)=\frac{1}{G^{-1}_R(\omega,k)+\kappa_+},
\end{eqnarray*}
where $G_R(\omega,k)$ is the two point function for $\mathcal{O}$ in the absence of the double trace deformation \cite{Iqbal:2011ae}. 

The double trace deformation displaces the poles of the Green's function and results in new zero modes. At the leading order of frequency one finds
\begin{eqnarray}\label{GHyb}
G^{(\kappa)}_R(\omega,k)=\frac{b_+^{(0)}(k)+b_-^{(0)}(k)\mathcal{G}_k(\omega)}{a_+^{(0)}(k)+b_+^{(0)}(k)\kappa_++\left(a_-^{(0)}(k)+b_-^{(0)}(k)\kappa_+\right)\mathcal{G}_k(\omega)}.
\end{eqnarray}
The poles of the deformed Green's function lie at $a_+^{(0)}(k)+b_+^{(0)}(k)\kappa_+=0$, which leads to
\begin{eqnarray*}
\kappa_c=3^{-\nu_U}\frac{\Gamma(-1-\Delta_+)\Gamma(-\frac{1+\Delta_-}{2}+\nu_k-\frac{ie}{3})\Gamma(-\frac{1+\Delta_-}{2}+\nu_k+\frac{ie}{3})}{\Gamma(-1-\Delta_-)\Gamma(-\frac{1+\Delta_+}{2}+\nu_k-\frac{ie}{3})\Gamma(-\frac{1+\Delta_+}{2}+\nu_k+\frac{ie}{3})}.
\end{eqnarray*}
Hybridized critical points correspond to $\kappa_+=\kappa_c$. Near this critical point, the Green's function at zero frequency and momentum limit takes the form
\begin{eqnarray*}
G^{(\kappa)}_R(\omega\rightarrow0,k\rightarrow0)=\frac{1}{\kappa_+-\kappa_c+k^2h_k+\omega h_\omega+\omega^2h_{\omega^2}+h C(\nu)(-i\omega)^{2\nu}}
\eeas
where
\beas
h_{k^2}=\frac{\partial_k^2\tilde{a}^{(0)}_+}{b^{(0)}_+}\bigg|_{\substack{k=0\\\kappa_+=\kappa_c}}
,\;\;\hspace*{0.75cm}h_{\omega}=\frac{\tilde{a}^{(0)}_+}{b^{(0)}_+}\bigg|_{\substack{k=0\\\kappa_+=\kappa_c}},\;\;\hspace*{0.75cm}h_{\omega^2}=\frac{\tilde{a}^{(2)}_+}{b^{(0)}_+}\bigg|_{\substack{k=0\\\kappa_+=\kappa_c}},
\;\;\hspace*{0.75cm}h=\frac{\tilde{a}^{(0)}_-}{b^{(0)}_+}\bigg|_{\substack{k=0\\\kappa_+=\kappa_c}}=\frac{-\nu}{\nu_U {b^{(0)}_+}^2},
\end{eqnarray*}
and $\mathcal{G}=C(\nu)(-i\omega)^{2\nu}$.
Since Eq. \eqref{UVleading} is parity invariant, there is no linear term of momentum in the above expansion and a similar reasoning leads to the absence of linear frequency term in the neutral case.

In the case of $\nu<1$, for a neutral scalar the $h_{\omega^2}$ term can be neglected and the pole of the Green's function originates from the IR region
\begin{eqnarray*}
\omega_*=i\left(\frac{\kappa_c-\kappa_+}{h C(\nu)}\right)^{\frac{1}{2\nu}}.
\end{eqnarray*}
In this case $hC(\nu)>0$. So for $\kappa_+\lessgtr\kappa_c$ we have Im$(\omega_*)\gtrless0$, which shows that unstable region corresponds to $\kappa_+<\kappa_c$.

\begin{figure}
\centering
\begin{subfigure}
\centering
\includegraphics[scale=.75]{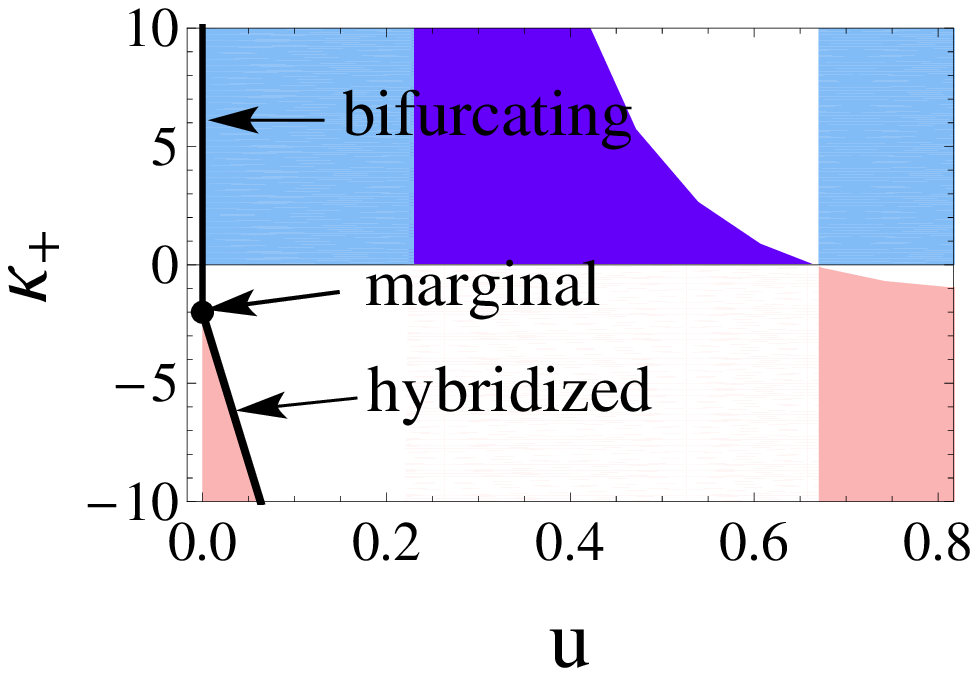}
\end{subfigure}
\begin{subfigure}
\centering
\includegraphics[scale=.71]{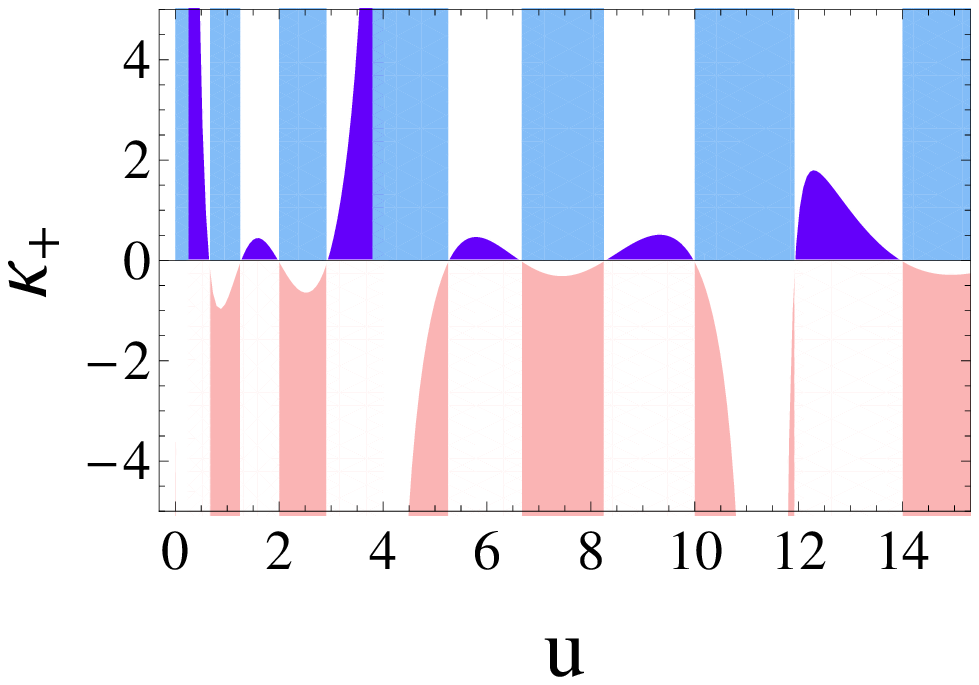}
\end{subfigure}
\caption{Phase diagrams for standard quantization of a neutral scalar. For $\kappa_+>0$, if there is no $\kappa_c>0$, the system is unstable (light blue regions). When $\kappa_c>0$ exists, the system is unstable for $0<\kappa_+<\kappa_c$ (dark blue regions). For $\kappa_+<0$, the system becomes unstable if $\kappa_c<0$ exists and $\kappa_+<\kappa_c$ (red regions). For $u<0$ the system is unstable because of the oscillatory region. The system is stable in the white regions.}
\label{K+2}
\end{figure}

The hybridized critical point is further investigated in Fig. \ref{K+2} where the critical value of $\kappa_{+}$ is plotted as a function of $u$. In the left plot, one can find different critical points in the standard quantization. The $u=0$ line refers to the bifurcating critical point while the border specified by $\kappa_c$ shows the hybridized critical point. The point of intersection corresponds to the marginal critical point. As it was shown, while $\kappa_+$ crosses down the $\kappa_+=\kappa_c$ curve, the pole of the Green's function crosses the origin of the complex $\omega$-plane causing an instability. This instability is guaranteed because $\kappa_c$ is a single valued function. Obviously this description is valid while our analytic calculation is valid in the low energy limit. The right plot is the same as the left one in a wider $u$ region where the zeroes of $\kappa_c$ could be seen. Note that in all plots of Fig. \ref{K+2} we have assumed the UV and IR instabilities are related. If they are completely unrelated, there is no stable region for $\kappa_+>0$ (see \cite{Gulotta:2010cu})\footnote{We thank Jie Ren for bringing our attention to this point.}.
\subsection{Marginal quantum critical point}

Approaching the bifurcating and hybridized quantum critical points simultaneously, a new critical behavior occurs known as marginal quantum critical point. More precisely one must consider both $u\to0$ and $\kappa_+\to\kappa_c$ limits at the same time. Since the IR scaling dimension is $\delta_k=\frac{1}{2}+\nu_k$, in the $\nu_k\to0$ limit the double trace operator becomes marginal. In the following we will read the retarded Green's function in this limit.

Consider a neutral scalar on the border of the oscillatory region $(m^2=-3)$, the solution in the inner region can be written in terms of Hankel functions. After considering ingoing boundary condition on the horizon one finds 
\begin{eqnarray*}
\phi_I(\xi)\sim \sqrt{\xi}\,H_0^{(1)}(\omega \xi).
\end{eqnarray*}
In terms of the UV region coordinates the above solution reads
\begin{eqnarray*}
\phi_I(\xi\rightarrow0)\sim \left[12(r-1)\right]^{-\frac{1}{2}}\left[\mathcal{G}_0+\ln12(r-1)\right],
\end{eqnarray*}
where $\mathcal{G}_0=-\gamma-\ln(\frac{-i\omega}{2})$. In the outer region the solution is
\begin{eqnarray*}
\phi_O(r)\sim \frac{1}{\sqrt{r^2-1}}\left(c_1+c_2 \ln\frac{r^2-1}{r^2+2}\right).
\end{eqnarray*}
By matching the IR and UV regions one can
read the retarded Green's function in this case as follows
\begin{eqnarray}\label{GM}
G=\frac{\ln\omega-2\ln6+6+\gamma-\frac{i\pi}{2}}{(\kappa_++2)(\ln\omega-2\ln6+\gamma-\frac{i\pi}{2})+6\kappa_+}.
\end{eqnarray}
The critical point occurs at $\kappa_+=-2$ (see Fig. \ref{K+2}). The logarithmic term in Eq. \eqref{GM} indicates the bahavior of marginal Fermi liquids which is discussed in \cite{Iqbal:2011aj}.
\section*{Conclusions}
In this paper we have studied holographic quantum critical points of a charged Lifshitz space-time. We have studied the near horizon geometry showing that it contains an AdS$_2$ factor (as in the RN black-brane case) which is dual to conformal quantum mechanics. We considered a charged scalar field in this background and worked out the $z$-dependence of the scaling dimension of the dual operator. In the case of $z=2$, we worked out the exact solution in the UV and IR regions and computed the analytic retarded Green's function in the leading order of frequency expansion. We have found the corresponding zero modes analytically and studied the stability regions of the model in the parameter space. We have also studied the bifurcating, hybridized and marginal quantum critical points of the system.

\vspace{5mm}
We can summarize the properties of the model as follows: 
\begin{itemize}
\item There are several exact information about different quantities in the model such as zero modes and the locus of superfluid surfaces.
\item This non-relativistic theory flows to an IR fixed point, but the dynamical exponent still affects the bosonic instabilities. The $z$ dependence of the IR scaling dimension of a scalar operator shows that strongly correlated systems described by larger $z$ are more stable under scalar condensation.
\item Stable IR geometry of this model supports standard quantization similar to the case of RN-AdS$_5$.
\item The system supports bifurcating phase transition where the generalized BF bound is violated in the IR geometry.  It also supports a hybridized QCP via adding double trace deformations to the model.
\item By studying these phase transitions one can not distinguish between relativistic and non-relativistic field theories in the UV region. 
\end{itemize}     
To further study this background one can probe it by fermionic fields and investigate the existence of Fermi surfaces and its QCPs. One can also study scalar instabilities in more general non-relativistic backgrounds, such as those with hyperscaling violation. 

\section*{Acknowledgments}
We would like to thank D. Allahbakhshi, J. Ren, A. Vahedi, and M. R. Tanhayi for useful discussions and also H. Ebrahim for her useful comments. We would also like to thank Prof. M. Alishahiha, for his useful comments and his supports during this work. 
\appendix

\section{IR Green's function}
In this appendix we compute the Green's function of a scalar field in the near horizon region, for both zero and finite temperature cases. These Green's functions in the RN-AdS black-brane case has been done in \cite{FILMV:2011}.
\subsection*{Zero temperature case}
We consider the action of a charged scalar minimally coupled to the gauge field as
\begin{eqnarray}\label{action}
S=- \int{d^{4}x \sqrt{-g}\left[g^{MN}\left(\partial_M +iqA_M\right)\phi^*\left(\partial_N -iqA_N\right)\phi+m^2\phi^* \phi\right]}
\end{eqnarray}
so the equation of motion after Fourier transforming like $\phi(r,x^\mu)=e^{-i\omega t+ik.x}\phi(r)$, becomes
\begin{eqnarray}\label{EOM1}
\Box \phi-\left(\frac{iq}{\sqrt{-g}}\partial_M\left(\sqrt{-g}A^M\right)+2iqA^M\partial_M+q^2A^2+m^2\right)\phi=0,
\end{eqnarray}
where the scalar d'Alembertian is defined as $\Box=\frac{1}{\sqrt{-g}}\partial_M\left(\sqrt{-g}\partial^M\right)$. 

Now consider a scalar field in the near horizon geometry of the Lifshitz space-time Eq. \eqref{AdS2}. Also notice that we only consider $A^{(2)}$ as the gauge field that can be coupled to the scalar. The equation of motion takes the form of
\begin{eqnarray}\label{EOM-AdS2}
\xi^2 \phi''+\left[\left(\omega \xi+e_z\right)^2-(m^2+k^2)l_z^2\right]\phi=0,
\end{eqnarray}
where $e_z=\frac{e}{z+1}=\frac{4q}{(z+1)\kappa}$. Near the boundary of AdS$_2$ (i.e. $\xi\to0$), the solution of the above equation becomes
\begin{eqnarray}\label{EOM-AdS2bdy}
\phi(\xi \rightarrow0)=\phi_0 \xi^{\frac{1}{2}-\nu_k}+\phi_1 \xi^{\frac{1}{2}+\nu_k},
\end{eqnarray}
where $\phi_0$ and $\phi_1$ are the source and the response functions respectively and 
\begin{eqnarray}\label{nu1}
\nu_k=\sqrt{\frac{1}{4}+(m^2+k^2)l_z^2-e_z^2}.
\end{eqnarray}
The conformal dimension of the dual operator is $\delta_k=\frac{1}{2}+\nu_k$. The near boundary solution Eq. \eqref{EOM-AdS2bdy}, near the matching region $(\xi \rightarrow 0)$, up to an overall factor takes the form of
\begin{eqnarray}\label{near matching}
\phi(\xi \rightarrow0) \simeq [12(r-1)]^{-\frac{1}{2}+\nu_k}+[12(r-1)]^{-\frac{1}{2}-\nu_k}\mathcal{G}_k.
\end{eqnarray}
Eq. \eqref{EOM-AdS2} could be solved exactly in terms of Whittaker functions as
\begin{eqnarray}\label{Whittaker}
\phi(\xi)=c_{in}W_{ie_z,\nu_k}(-2i\omega \xi)+c_{out}W_{-ie_z,\nu_k}(2i\omega \xi),
\end{eqnarray}
where applying the ingoing boundary condition at the horizon leads to $c_{out}=0$.
Expanding this solution near the boundary of the AdS$_2$ region, one can read the IR Green's function as
\begin{eqnarray*}
\mathcal{G}_k=e^{-i\pi \nu_k}\frac{\Gamma(-2\nu_k)}{\Gamma(2\nu_k)}\frac{\Gamma(\frac{1}{2}-ie_z+\nu_k)}{\Gamma(\frac{1}{2}-ie_z-\nu_k)}(2\omega)^{2\nu_k}.
\end{eqnarray*}
As in the relativistic case, the IR Green's function scales with frequency for arbitrary momenta.

\subsection*{Finite temperature case}
In this case we consider the following metric as the near horizon limit at finite temprature of the Lifshitz black-brane
\begin{eqnarray*}
ds^2=\frac{l_z^2}{\xi^2}\left(-\left(1-\frac{\xi^2}{\xi_h^2}\right)d\tau^2+\frac{d\xi^2}{\left(1-\frac{\xi^2}{\xi_h^2}\right)}\right)+dx_2^2,
\eeas
where the gauge field is
\beas
A^{(2)}_\tau=\frac{4}{(z+1)\kappa\xi}\left(1-\frac{\xi}{\xi_h}\right)
\end{eqnarray*}
and its Hawking temperature\footnote{If one does not set the horizon of the whole geometry to unity, the last term would have been $r_*^2dx_2^2$ where $r_*=\left(\frac{z}{z+2}\right)^{\frac{1}{2z+2}}r_0$ is the scale of spatial momenta.} reads $T=(2\pi\xi_h)^{-1}$. The equation of motion for a scalar field in this background is
\begin{eqnarray*}
\phi''+\frac{2\xi}{\xi^2-\xi_h^2}\phi'+\left[\frac{\left(\omega+e_z\left(\frac{1}{\xi}-\frac{1}{\xi_h}\right)\right)^2}{\left(1-\frac{\xi^2}{\xi_h^2}\right)^2}-\frac{m^2+k^2}{\xi^2-\frac{\xi^4}{\xi_h^2}}l_z^2\right]\phi=0,
\end{eqnarray*}
which its solution is
\begin{eqnarray*}
\phi(\xi)=\left(\frac{\xi+\xi_h}{\xi-\xi_h}\right)^{\frac{i\omega \xi_h}{2}}\bigg[c_1 \left(1+\frac{\xi_h}{\xi}\right)^{\nu_k-\frac{1}{2}}F\left(\frac{1}{2}-\nu_k-ie_z,\frac{1}{2}-\nu_k+ie_z-i\omega\xi_h,1-2\nu_k,\frac{2\xi}{\xi +\xi_h}\right)+\nonumber\\
(-2)^{2\nu}c_2 \left(1+\frac{\xi_h}{\xi}\right)^{-\nu_k-\frac{1}{2}}F\left(\frac{1}{2}+\nu_k-ie_z,\frac{1}{2}+\nu_k+ie_z-i\omega\xi_h,1+2\nu_k,\frac{2\xi}{\xi +\xi_h}\right)\bigg],
\end{eqnarray*}
and $\nu_k$ was defined in \eqref{nu1}. The ingoing boundary condition leads to
\begin{eqnarray*}
c_2=(-1)^{-2\nu}\frac{\Gamma(-2\nu_k)}{\Gamma(2\nu_k)}\frac{\Gamma(\frac{1}{2}-ie_z+\nu_k)\Gamma(\frac{1}{2}+ie_z+\nu_k-i\omega\xi_h)}{\Gamma(\frac{1}{2}-ie_z-\nu_k)\Gamma(\frac{1}{2}+ie_z-\nu_k-i\omega\xi_h)}c_1,
\end{eqnarray*}
and the retarded Green's function takes the form of
\begin{eqnarray}\label{scalarIRGreenTemp}
\mathcal{G}_k^T(\omega)=(4\pi T)^{2\nu}\frac{\Gamma(-2\nu_k)}{\Gamma(2\nu_k)}\frac{\Gamma(\frac{1}{2}-ie_z+\nu_k)\Gamma(\frac{1}{2}+ie_z+\nu_k-\frac{i\omega}{2\pi T})}{\Gamma(\frac{1}{2}-ie_z-\nu_k)\Gamma(\frac{1}{2}+ie_z-\nu_k-\frac{i\omega}{2\pi T})}.
\end{eqnarray}

\end{document}